\documentclass{ws-ijmpd}

\begin{document}

\markboth{U. Mukhopadhyay, S. Ray \& F. Rahaman} {Dark Energy
Models With Variable Equation Of State Parameter}

%
\catchline{}{}{}{}{}
%

\title{Dark Energy Models With Variable Equation Of State Parameter}

\author{Utpal Mukhopadhyay}
\address{Satyabharati Vidyapith, Barasat, North 24 Parganas,
Kolkata 700 126, West Bengal, India\\ umsbv@yahoo.in}

\author{Saibal Ray}
\address{Department of Physics, Government College of Engineering \& Ceramic Technology, Kolkata 700 010, West Bengal, India\\ saibal@iucaa.ernet.in}

\author{Farook Rahaman}
\address{Department of Mathematics, Jadavpur University,
  Kolkata 700 032, West Bengal, India\\  farook\_rahaman@yahoo.com}

\maketitle

\begin{history}
\received{Day Month Year} \revised{Day Month Year}
\end{history}

\begin{abstract}
Two phenomenological variable $\Lambda$ models, viz. $\Lambda \sim
(\dot a/a)^2$ and $\Lambda \sim \rho$ have been studied under the
assumption that the equation of state parameter $\omega$ is a
function of time. The selected $\Lambda$ models are found to be
equivalent both in four and five dimensions. The possibility of
signature flip of the deceleration parameter is also shown.

\end{abstract}

\section{Introduction}
Recent years have witnessed the emergence of the idea of an
accelerating Universe and due to some observational results
\cite{Dunlop1996,Spinard1997,Riess1998,Perlmutter1999}, it is now
established that the Universe is accelerating. This signifies a
paradigm shift in cosmological research from {\it expanding
Universe} to {\it accelerated expanding Universe}. Now, the
problem lies in detecting an exotic type of unknown repulsive
force, termed as dark energy, which is driving this acceleration.
A key factor in dark energy investigation is the equation of state
parameter $\omega$, which relates pressure and density through an
equation of state of the form $p=\omega\rho$. Due to lack of
observational evidence in making a distinction between constant
and variable $\omega$, usually the equation of state parameter is
considered as a constant \cite{Kujat2002,Bartelmann2005} with
phase wise values $0$, $1/3$, $-1$ and $+1$ for dust, radiation,
vacuum fluid and stiff fluid dominated Universe respectively. But
in general, $\omega$ is a function of time or redshift
\cite{Chevron2000,Zhuravlev2001,Peebles2003,Jimenez2003,Das2005}.
For instance, quintessence models involving scalar fields give
rise to time-dependent $\omega$
\cite{Ratra1988,Turner1997,Caldwell1998,Liddle1999,Steinhardt1999}.
There is literature available on models with varying fields, such
as cosmological model with a time dependent equation of state in a
Kaluza-Klein metric, cosmological model with a viscous fluid in a
Kaluza Klein metric and wormholes with varying equation of state
parameter\cite{Bhui2005,Rahaman2006,Rahaman2009}. So, there are
enough grounds for considering $\omega$ as time-dependent for a
better understanding of the cosmic evolution.

Now, various types of physical models including phenomenological
ones (with time-dependent $\Lambda$) are contemplated for
unveiling the nature of dark energy. It may be mentioned here that
phenomenological time-varying $\Lambda$ models were proposed
initially even before the emergence of the idea of cosmic
acceleration for solving the well known cosmological constant
problem \cite{Freese1987,Ozer1987,Chen1990,Carvalho1992,Lima1994}.
At present, various kinematical $\Lambda$ models of
phenomenological character are proposed for explaining the present
acceleration. In a recent work, Ray et al. \cite{Ray2007} have
shown the equivalence of three phenomenological $\Lambda$ models,
viz. $\Lambda \sim (\dot a/a)^2$, $\Lambda \sim \ddot a/a$ and
$\Lambda \sim \rho$ in four dimensional space-time for constant
$\omega$ while the same three models of $\Lambda$ were shown to be
equivalent in five dimensional Kaluza-Klein type model by Pradhan
et al.\cite{Pradhan2005} treating $\omega$ as a constant quantity.
So, it is quite natural to investigate the behaviour of the above
three models when the equation of state parameter is a function of
time. In fact, Mukhopadhyay et al.
\cite{Mukhopadhyay2008,Mukhopadhyay2009} have already made an
attempt in this direction by investigating two other dynamical
$\Lambda$ models, viz. $\dot \Lambda\sim H^3$ and $\Lambda\sim\dot
H$ with variable equation of state parameter.

In the present work, an investigation of the models $\Lambda \sim
(\dot a/a)^2$ and $\Lambda \sim \rho$ are made under the {\it
ansatz} $\omega=\omega(t)$ both in four and five dimensional
space-time. For a flat Universe, these two models are shown to be
equivalent for variable $\omega$ both in four and five dimensions.
Some other features, revealed by letting $\omega$ be a function of
time, have also been observed.

\section{Field Equations in four dimensions and their solutions}
The Einstein field equations are given by
\begin{eqnarray}
R^{ij}-\frac{1}{2}Rg^{ij}= -8\pi G\left[T^{ij}-\frac{\Lambda}{8\pi
G}g^{ij}\right],
\end{eqnarray}
where the cosmological term $\Lambda$ is time-dependent, i.e.
$\Lambda = \Lambda(t)$ and $c$, the velocity of light in vacuum,
is assumed to be unity.

Let us consider the Robertson-Walker metric
\begin{eqnarray}
ds^2=-dt^2+a(t)^2\left[\frac{dr^2}{1-kr^2}+r^2(d\theta^2+sin^2\theta
 d\phi^2)\right],
\end{eqnarray}
where $k$, the curvature constant, assumes the values $-1$, $0$
and $+1$ for open, flat and closed models of the Universe
respectively and $a=a(t)$ is the scale factor. For a flat Universe
($k=0$) and the spherically symmetric metric (2), field
equations~(1) yield Friedmann and Raychaudhuri equations
respectively, given by
\begin{eqnarray}
3H^2+\frac{3k}{a^2}= 8\pi G\rho+\Lambda,
\end{eqnarray}
\begin{eqnarray}
3H^2+3\dot H= -4\pi G(\rho+3p)+\Lambda,
\end{eqnarray}
where $\rho$ and $p$ are the cosmic matter-energy density and
pressure respectively and the Hubble parameter $H$ is related to
the scale factor by $H=\dot a/a$.

Let us now choose the barotropic equation of state
\begin{eqnarray}
p= \omega\rho.
\end{eqnarray}
Here we assume that the equation of state parameter $\omega$ is
time-dependent i.e. $\omega=\omega(t)$ such that $\omega =
(t/\tau)^n-1$, where $\tau$ is a constant having dimension of
time.

From equation (3), for flat Universe, we get
\begin{eqnarray}
\rho = \frac{3H^2-\Lambda}{8\pi G}.
\end{eqnarray}
Using equation (5) and (6) in (4) one can get, after some
manipulation, the following differential equation
\begin{eqnarray}
\frac{dH}{dt} = \frac{(1+\omega)(\Lambda-3H^2)}{2}.
\end{eqnarray}

\subsection{$\Lambda=3\alpha H^2$} Let us use the {\it ansatz}
$\Lambda=3\alpha H^2$ (where $\alpha$ is a free parameter) along
with $\omega = (t/\tau)^n-1$. Then equation (7) reduces to
\begin{eqnarray}
\frac{dH}{H^2} = -\frac{3(1-\alpha)t^n}{2\tau^n}dt.
\end{eqnarray}
Solving equation (8), we get
\begin{eqnarray}
H = \frac{2(n+1)\tau^n}{3(1-\alpha)t^{n+1}}.
\end{eqnarray}
Writing $H=\dot a/a$ in equation (9) and integrating it further we
get our solution set as
\begin{eqnarray}
a(t) = C_1 e^{
 \frac{-2(n+1)}{3n(1-\alpha)}(t/\tau)^{-n}},
\end{eqnarray}
\begin{eqnarray}
\rho(t) = \frac{(n+1)^2\tau^{2n}}{6\pi
 G(1-\alpha)}t^{-2(n+1)},
\end{eqnarray}
\begin{eqnarray}
\Lambda(t)=
 \frac{4\alpha(n+1)^2\tau^{2n}}{3(1-\alpha)^2}t^{-2(n+1)},
\end{eqnarray}
where $C_1$ is an integration constant.

It is interesting to note that putting $n=0$ (i.e. $\omega=0$) and
{\bf$\tau=1$} in equations (9), (11) and (12) we can recover the
same expressions for $H(t)$, $\rho(t)$ and $\Lambda(t)$
respectively as those obtained by Ray et al. \cite{Ray2007}. But,
equation (10) indicates that $n$ cannot be equal to zero.
Moreover, the expression for $a(t)$ in equation (10) differs from
that of  Ray et al. \cite{Ray2007}. So, in general, the constant
$\omega$ of Ray et al. \cite{Ray2007} and time-varying $\omega$ of
this case present us with different situations.

By using equations (6) and (9), we get \cite{Ray2007}
\begin{eqnarray}
\alpha = 1-\Omega_m = \Omega_{\Lambda},
\end{eqnarray}
where, in absence of any curvature, matter density $\Omega_m$ and
dark energy density $\Omega_{\Lambda}$ are related by the equation
\begin{eqnarray}
\Omega_m + \Omega_{\Lambda} = 1.
\end{eqnarray}

\subsection{$\Lambda=8\pi G\gamma\rho$} Using the {\it ansatz}
$\Lambda=8\pi G\gamma\rho$ (where $\gamma$ is a parameter) in
equation (7) we get the differential equation
\begin{eqnarray}
\frac{dH}{dt} = -\frac{3H^2}{2(\gamma+1)\tau^n}t^n.
\end{eqnarray}
Solving equation (15), we get
\begin{eqnarray}
H = \frac{\dot a}{a}=\frac{2(\gamma+1)(n+1)\tau^n}{3t^{n+1}}.
\end{eqnarray}
Again, solving equation (16) we get our solution set as
\begin{eqnarray}
a(t) = C_2 e^{\frac{-2(n+1)(\gamma+1)}{3n}(t/\tau)^{-n}},
\end{eqnarray}
\begin{eqnarray}
\rho(t)= \frac{(\gamma+1)(n+1)^2\tau^{2n}}{6\pi
 G}t^{-2(n+1)},
\end{eqnarray}
\begin{eqnarray}
\Lambda(t)
=\frac{4\gamma(\gamma+1)(n+1)^2\tau^{2n}}{3}t^{-2(n+1)},
\end{eqnarray}
where $C_2$ is an integration constant. In this case also, for
 $n=0$ and $\tau=1$ we get back the same expressions for $H(t)$,
$\rho(t)$ and $\Lambda(t)$ as those obtained by Ray et
al.~\cite{Ray2007}. But here again $n=0$ is forbidden via equation
(17). Thus for the $\Lambda\sim\rho$ case also, variable $\omega$
and constant $\omega$ present us with different situations.

\section{Field Equations in five dimensions and their solutions}
Kaluza-Klein type Robertson-Walker metric is given by
\begin{eqnarray}
ds^2=
-dt^2+a(t)^2\left[\frac{dr^2}{1-kr^2}+r^2(d\theta^2+sin^2\theta
d\phi^2)+(1-kr^2)d\psi^2\right].
\end{eqnarray}
For the metric (20), the field equations (1) yield the following
two differential equations
\begin{eqnarray}
6\left(H^2+\frac{k}{a^2}\right) = 8\pi\rho + \Lambda(t),
\end{eqnarray}
\begin{eqnarray}
3\dot H+6H^2+3\frac{k}{a^2} = -8\pi p+\Lambda(t),
\end{eqnarray}
where $G=c=1$.

From (21), for flat Universe ($k=0$), we can write
\begin{eqnarray}
\rho = \frac{6H^2-\Lambda}{8\pi}.
\end{eqnarray}
Using equations (5) and (23), for $k=0$, we obtain
\begin{eqnarray}
3\dot H+6(1+\omega)H^2 = (1+\omega)\Lambda.
\end{eqnarray}

\subsection{$\Lambda=3\alpha H^2$}
Let us substitute $\omega= (t/\tau)^n-1$ in equation (24), where
$\tau$ has the same significance as before. Then from equation
(24), after use of the {\it ansatz} $\Lambda=3\alpha H^2$, we get
\begin{eqnarray}
\dot H = (\alpha-2)(t/\tau)^n H^2.
\end{eqnarray}
Solving equation (26), we have
\begin{eqnarray}
H = \frac{(n+1)\tau^n}{(2-\alpha)t^{n+1}}.
\end{eqnarray}
Again, writing $H=\dot a/a$ in equation (26) and solving the
resulting differential equation we get our solution set as
\begin{eqnarray}
a(t) = C_3 e^{\frac{-(n+1)}{n(2-\alpha)}(t/\tau)^{-n}},
\end{eqnarray}
\begin{eqnarray}
\rho(t) = \frac{3(n+1)^2\tau^{2n}}{8\pi
 (2-\alpha)}t^{-2(n+1)},
\end{eqnarray}
\begin{eqnarray}
\Lambda(t)=
 \frac{3\alpha(n+1)^2\tau^{2n}}{(2-\alpha)^2}t^{-2(n+1)},
\end{eqnarray}
where $C_3$ is an integration constant.

Equation (27) shows that for physical validity $n\neq 0$. Also,
 from equation (26) we find that $\alpha<2$ if $\tau$ is positive.
But, equation (29) tells us that for a repulsive $\Lambda$,
$\alpha>0$. So combining these two cases we can write
$0<\alpha<2$.

\subsection{$\Lambda= 8\pi\gamma\rho$} Using the {\it ansatz}
$\Lambda= 8\pi\gamma\rho$, equation (24) reduces to
\begin{eqnarray}
\dot H = -\frac{2t^nH^2}{(\gamma+1)\tau^n}.
\end{eqnarray}
Solving equation (30), we get
\begin{eqnarray}
H = \frac{(n+1)(\gamma+1)\tau^n}{2t^{n+1}}.
\end{eqnarray}
Writing $H=\dot a/a$ in equation (31) and solving the resulting
differential equation, we get our solution set as
\begin{eqnarray}
a(t) = C_4 e^{\frac{-(n+1)(\gamma+1)}{2n}(t/\tau)^{-n}},
\end{eqnarray}
\begin{eqnarray}
\rho(t) =
 \frac{3(n+1)^2(\gamma+1)\tau^{2n}}{16\pi} t^{-2(n+1)},
\end{eqnarray}
\begin{eqnarray}
\Lambda(t) =
 \frac{3\gamma(\gamma+1)(n+1)^2\tau^{2n}}{2} t^{-2(n+1)},
\end{eqnarray}
where $C_4$ is an integration constant.

 Equation (32) shows that, for physical validity $n\neq 0$.
 Again, by equation (31), $\gamma>-1$ for a positive $\tau$, whereas equation (34)
 tells us that for repulsive $\Lambda$ either $\gamma>0$ or $\gamma<-1$.
 So, a positive $\gamma$ will make all the equations (31), (33) and (34)
physically valid.

\section{Equivalence of $\Lambda$ models}
\subsection{Four dimensional case}
Using equations (6) and (9) it is easy to obtain \cite{Ray2007}
\begin{eqnarray}
\alpha = 1-\Omega_m = \Omega_{\Lambda},
\end{eqnarray}

where, in the absence of any curvature, matter-energy density
$\Omega_m$ ($=8\pi G \rho/3H^2$) and dark energy density $
\Omega_{\Lambda}$ ($=\Lambda/3H^2$) are related by the equation
\begin{eqnarray}
\Omega_m + \Omega_{\Lambda} = 1.
\end{eqnarray}
Similarly, using equation (16) and (18), we find that $\gamma$ is
related to $\Omega_m$ and $\Omega_{\Lambda}$ through the relation
\begin{eqnarray}
\gamma = \frac{\Omega_{\Lambda}}{\Omega_m}.
\end{eqnarray}
Now, by the use of equations (35) and (37) we find that equations
(10) and (17) differ by a constant while the equations (11) and
(12) become identical to equations (18) and (19). This means that,
$\Lambda\sim(\dot a/a)^2$ and $\Lambda\sim\rho$ models are
equivalent.

\subsection{Five dimensional case}
Coming to the five dimensional Kaluza-Klein model, we find that
with the help of equations (23) and (26), one can easily obtain
\begin{eqnarray}
\alpha = 1+\Omega_{\Lambda},
\end{eqnarray}
for $\Lambda\sim(\dot a/a)^2$ model. It is to note that equation
(36) still holds in the 5D case.

Similarly, using equations (31) and (33), we get
\begin{eqnarray}
\gamma = \frac{1+\Omega_{\Lambda}}{\Omega_m},
\end{eqnarray}
for $\Lambda\sim\rho$ model.

With the help of equations (38) and (39) it is easy to show that
the equations (27) and (32) differ by a constant while equations
(29) and (30) become identical to the equations (33) and (34)
respectively. This implies that $\Lambda\sim(\dot a/a)^2$ and
$\Lambda\sim\rho$ models are equivalent for five dimensional
space-time also.

\section{Physical Features of the models}

\subsection{Some Features of $\tau$ and $\omega$} Apart from being
a mere mathematical parameter, $\tau$ bears some deeper physical
significance also. In some works
\cite{Dymnikova1998,Dymnikova2000,Dymnikova2001} it has been shown
that $\tau$ represents the time-scale of evaporation of
Bose-Einstein condensates which include a time-dependent
$\Lambda$. In the present work, $\tau$ is linked with the equation
of state parameter $\omega$ and cosmic time $t$. The exact nature
of $\omega$ and $t$ dependence of $\tau$ may be explored elsewhere
after $n$ is specified properly.

Although in the present work the equation of state parameter
$\omega$ is taken as time-dependent, it can be a function of the
red-shift $z$ or scale factor $a$ as well. The red-shift
dependence of $\omega$ can be linear like $\omega (z)=
\omega_0+{\omega} {\prime} z$  with ${\omega} {\prime}= (d{\omega}
{\prime} /dz)_z=0$~\cite{Huterer2001,Weller2002} or non-linear as
$\omega (z)= \omega_0+\omega_1
z/(1+z)$~\cite{Polarski2001,Linder2003}. Now, in the present work,
the structure of $\omega$ suggests that it is of the form $\omega
(t)= \omega_0+\omega_1 t^n$ which can be regarded as a
generalization of the special form $\omega (t)= \omega_0+\omega_1
t$ \cite{Mukhopadhyay2007}.

Now, $\tau$ can be expressed as
\begin{eqnarray}
\tau = \frac{t}{(1+\omega)^{1/n}}.
\end{eqnarray}
Since $\tau$ has dimensions of time, it cannot be negative. This
implies that $\omega>-1$ for integral values of $n$. SN Ia data
suggest that $-1.67<\omega<-0.62$ \cite{Knop2003} while the limit
imposed on $\omega$ by a combination of SN Ia data (with CMB
anisotropy) and galaxy clustering statistics \cite{Tegmark2004} is
$-1.33<\omega<-0.79$ . So, if the present work is compared with
the above-mentioned experimental results then, one can conclude
that the limit of $\omega$ provided by the equation (40) may be
accommodated within the acceptable range (Figs 1 and 2). Also it
is clear that for the present dust-filled Universe ($\omega=0$),
$\tau$ is equal to $t$ whereas for vacuum fluid $\tau$ becomes
meaningless.

\begin{figure*}
\begin{center}
\vspace{0.5cm}
\includegraphics[width=0.8\textwidth]{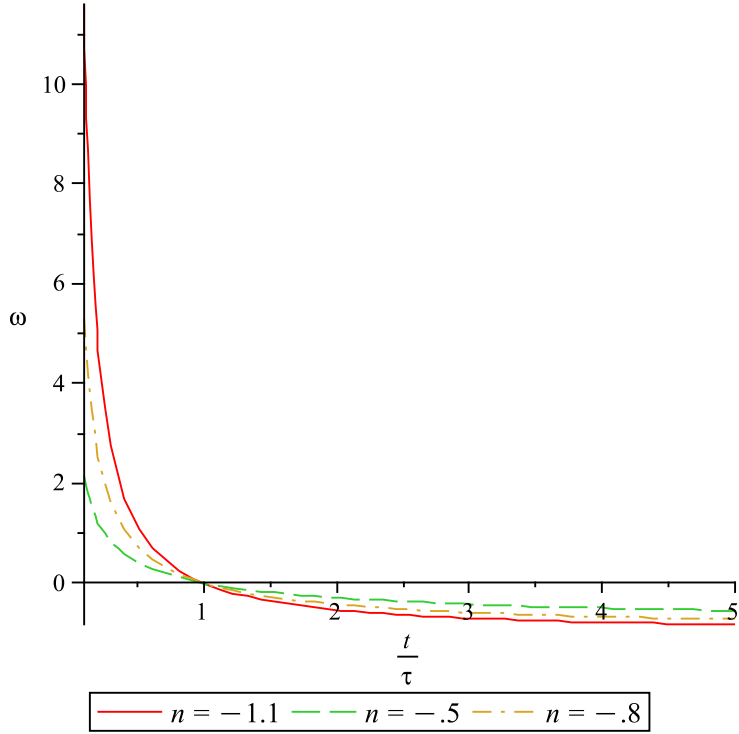}
\caption{The plot of $\omega$  vs. t for different values of n in
early epoch.} \label{fig:451}
\end{center}
\end{figure*}

\begin{figure*}
\begin{center}
\vspace{0.5cm}
\includegraphics[width=0.8\textwidth]{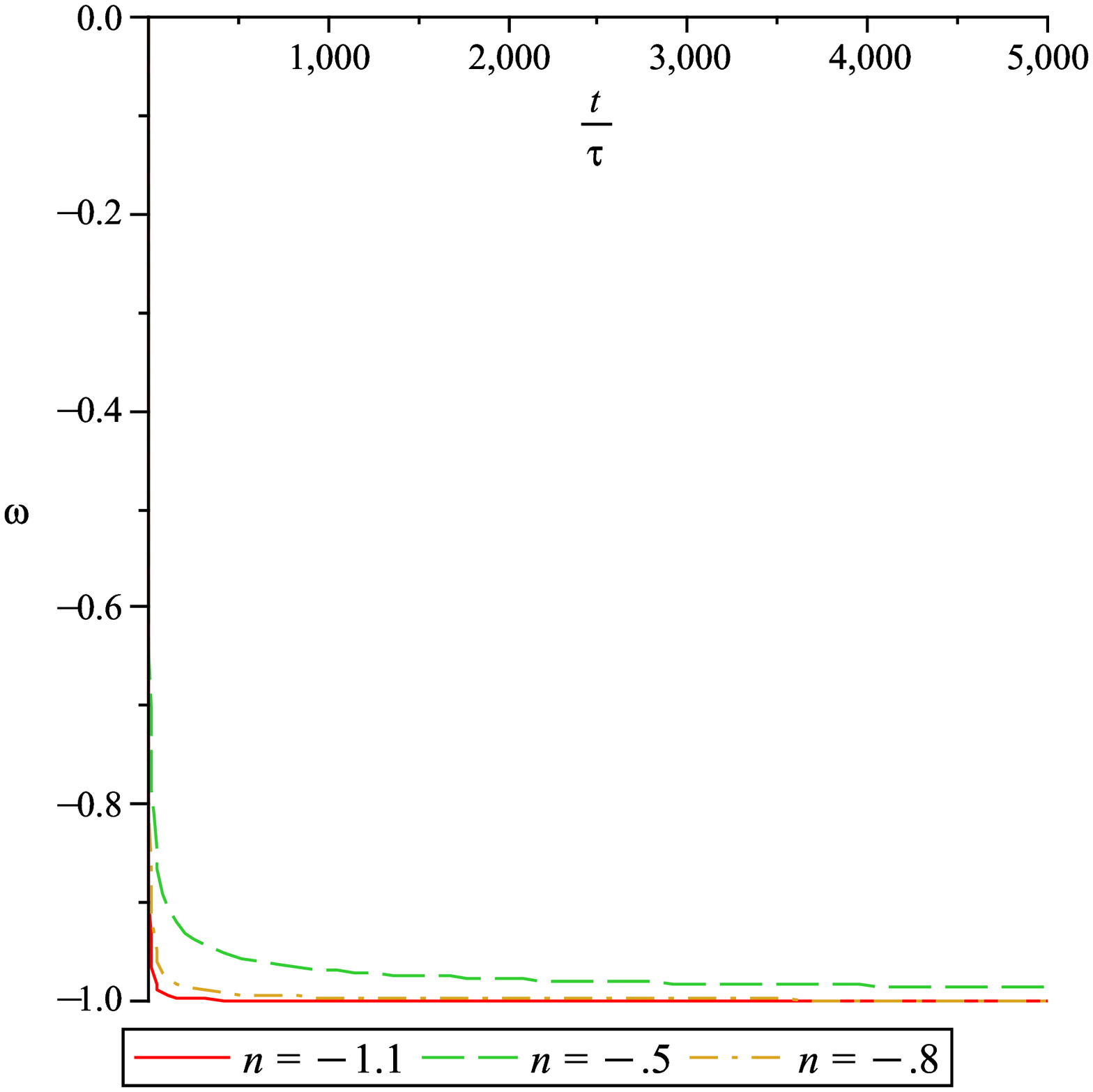}
\caption{The plot of $\omega$  vs. t for different values of n in
late time.} \label{fig:452}
\end{center}
\end{figure*}

\subsection{Calculation of the deceleration parameter} The deceleration
parameter $q$ is a very important factor for understanding cosmic
evolution. Particularly, after the emergence of the idea of an
accelerating Universe, the role of this parameter has become even
more important. This is because some recent works
\cite{Riess2001,Amendola2003,Padmanabhan2003} have demonstrated
that the present acceleration is a phenomenon of the recent past
and was preceded by a decelerating phase. This cosmological
picture supports the presently favored $\Lambda$-CDM Universe. So,
during cosmic evolution the deceleration parameter $q$ must have
undergone a change of sign from a positive to a negative value.
So, calculation of $q$ from our model is essential for detecting a
possible signature flip of $q$.

Now, using equations (9) and (35) we can obtain
\begin{eqnarray}
q = -\left[1+\frac{\dot H}{H^2}\right] = -[1-1.5\Omega_m
(t/\tau)^n].
\end{eqnarray}
Equation (42) shows that in the four dimensional case the
expression for $q$ contains a time factor and hence with a
suitable choice of $n$, $q$ can show its change of sign (Figs 3
and 4). Similarly, in the five dimensional case, using equations
(26) and (38) the expression for $q$ reduces to
\begin{eqnarray}
q = -[1-\Omega_m (t/\tau)^n].
\end{eqnarray}
Equation (42) shows that $q$ is time-dependent for the five
dimensional case also and hence may change its sign during cosmic
evolution (Figs 5 and 6). So, for $\Lambda\sim(\dot a/a)^2$ model,
$q$ shows the possibility of a change of sign in both four and
five dimensions. Since $\Lambda\sim(\dot a/a)^2$ and
$\Lambda\sim\rho$ models are equivalent, it is clear that
$\Lambda\sim\rho$ model will also show change of sign of $q$.

Again, $q$ can be written in terms of $\omega$ as
\begin{eqnarray}
q = -[1-1.5\Omega_m(1+\omega)],
\end{eqnarray}
in the four dimensional case and
\begin{eqnarray}
q = -[1-\Omega_m(1+\omega)],
\end{eqnarray}
in the five dimensional case.

\begin{figure*}
\begin{center}
\vspace{0.5cm}
\includegraphics[width=0.8\textwidth]{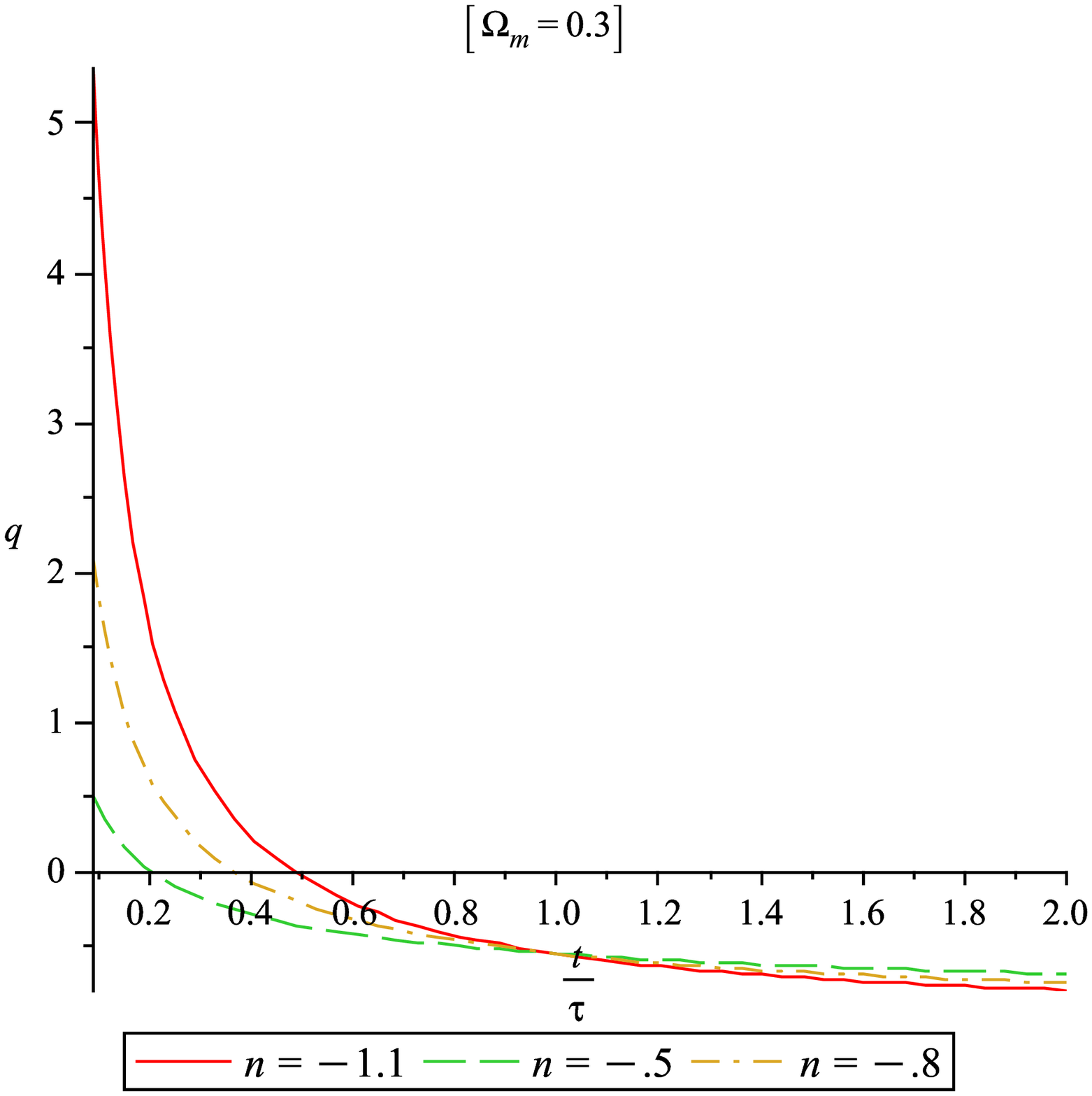}
\caption{The plot of q vs. t for different values of n with fixed
$\Omega_m$ in four dimension.} \label{fig:41}
\end{center}
\end{figure*}

\begin{figure*}
\begin{center}
\vspace{0.5cm}
\includegraphics[width=0.8\textwidth]{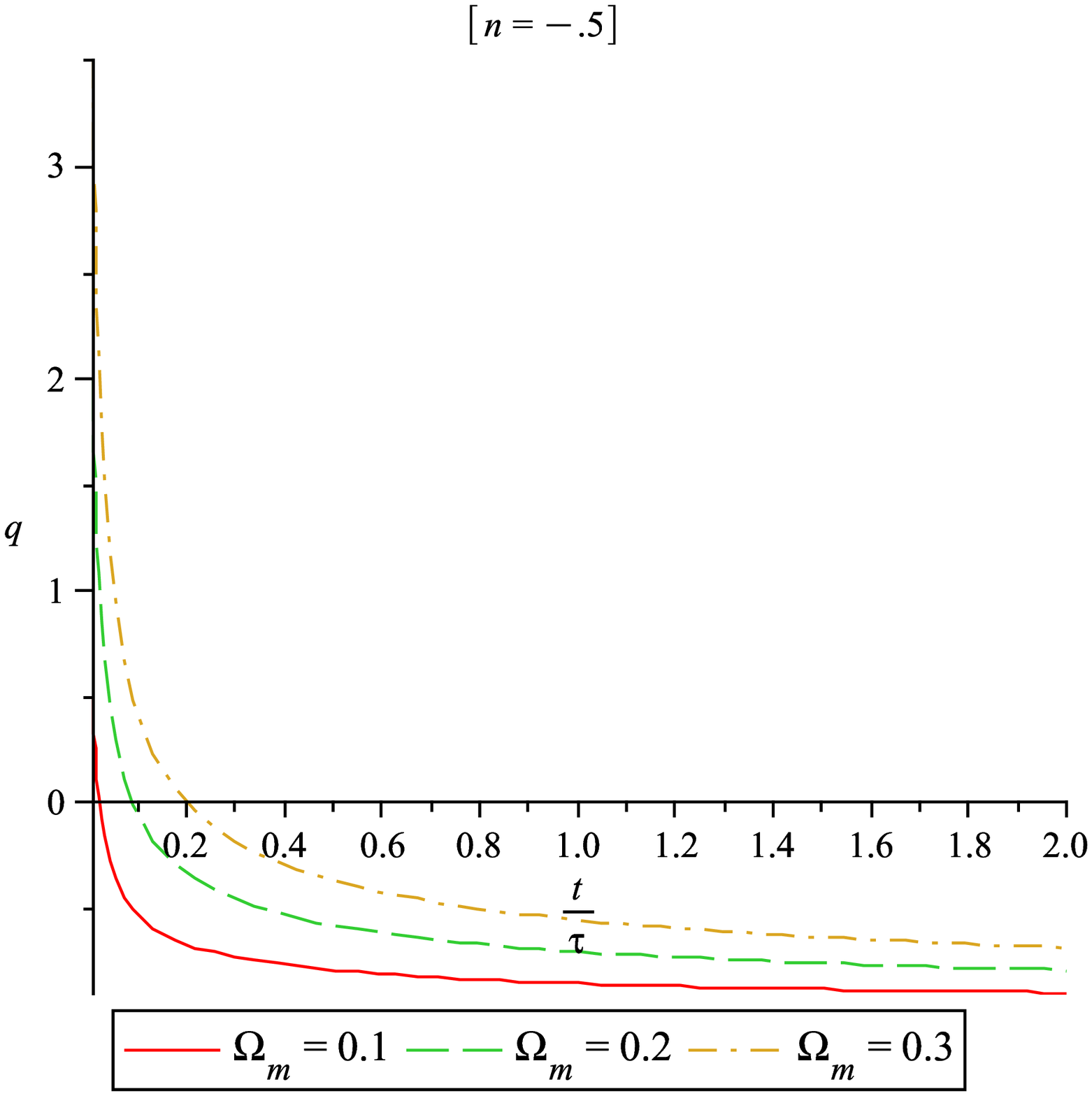}
\caption{The plot of q vs. t for different values of $\Omega_m$
with fixed n in four dimension.} \label{fig:42}
\end{center}
\end{figure*}

\begin{figure*}
\begin{center}
\vspace{0.5cm}
\includegraphics[width=0.8\textwidth]{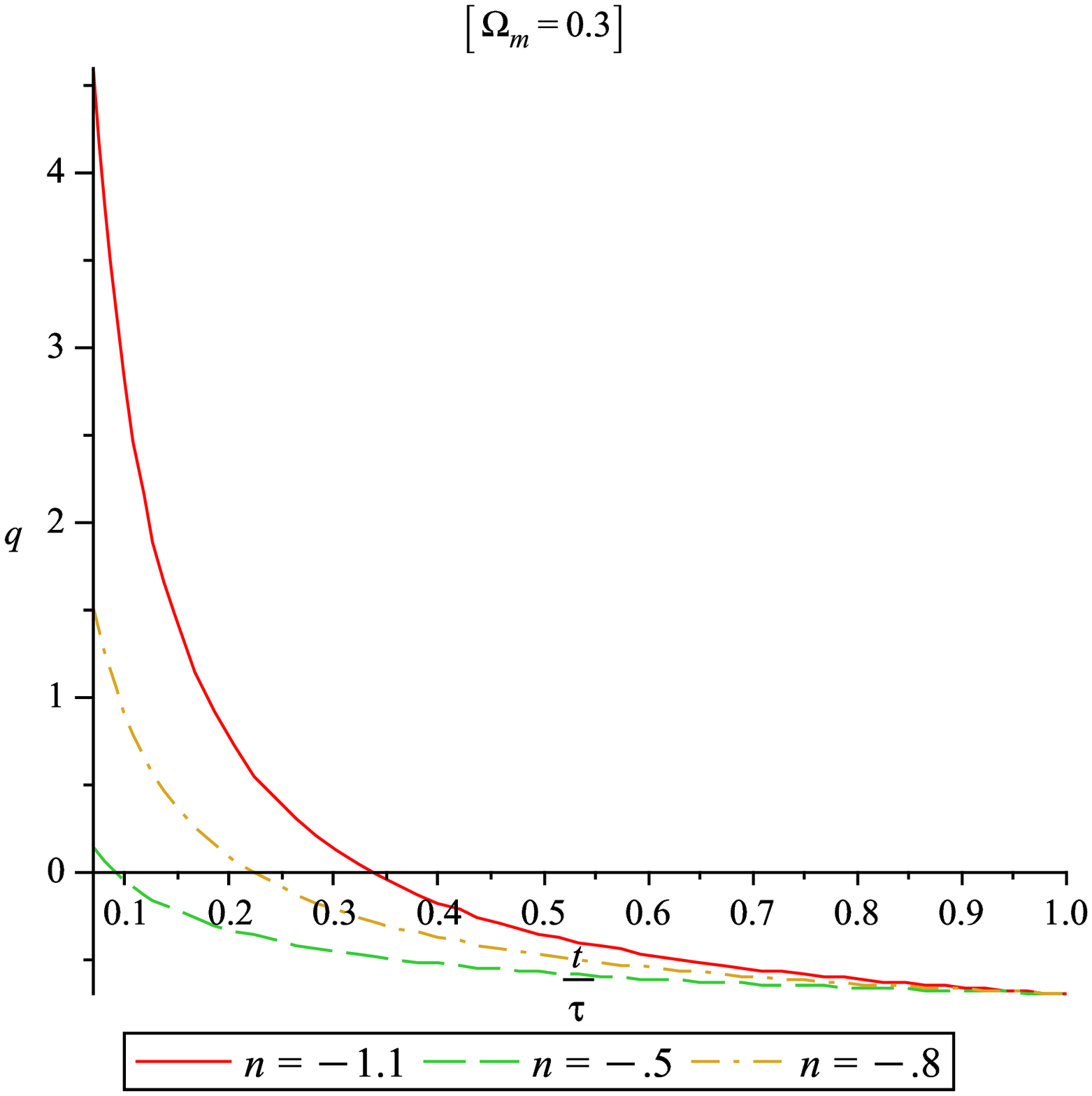}
\caption{The plot of q vs. t for different values of n with fixed
$\Omega_m$ in five dimension.} \label{fig:51}
\end{center}
\end{figure*}

\begin{figure*}
\begin{center}
\vspace{0.5cm}
\includegraphics[width=0.8\textwidth]{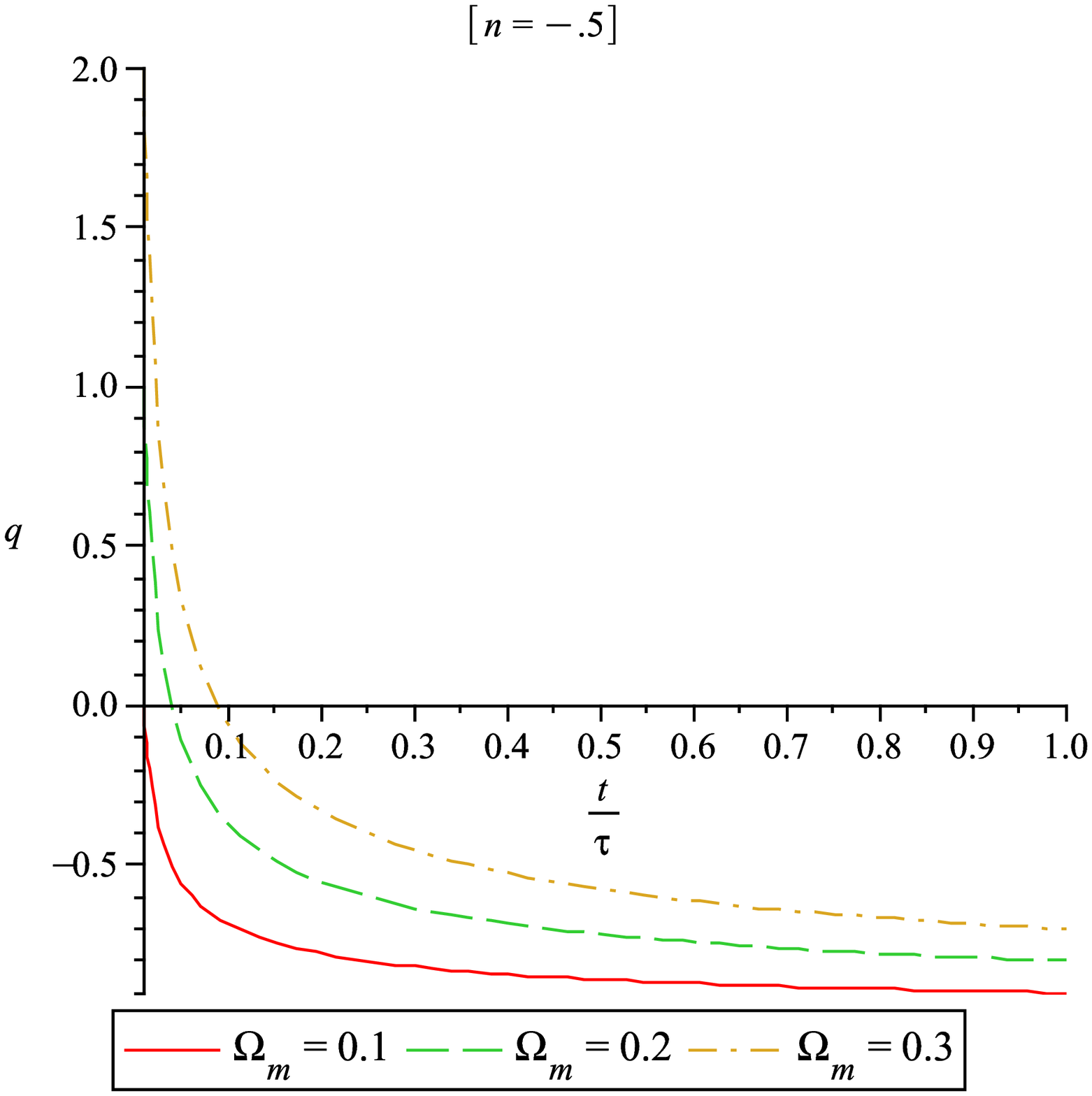}
\caption{The plot of q vs. t for different values of $\Omega_m$
with fixed n in five dimension.} \label{fig:52}
\end{center}
\end{figure*}

Now, considering various experimental and theoretical results, one
can safely assume that the present value of the cosmic
matter-energy density ($\Omega_{m0}$)as $0.33$ \cite{Ray2007}
(though seems rather a bit high value as given WMAP5 results). So,
for the present dust-filled Universe ($\omega=0$), in the four
dimensional case the value of $q$ comes out as $-0.505$ which is
in nice agreement with the present accepted range of $q$ for an
accelerating Universe \cite{Sahni1999}. For the five dimensional
case, $q=-0.67$ for the same set of values of $\Omega_{m0}$ and
$\Omega$ and hence that also supports the idea of an accelerating
Universe.

\section{Discussion}
Although some specific constant values for the equation of state
parameter $\omega$ are used for different phases of the cosmic
evolution, that technique suffers from some kind of discreteness
problem. The transition from one cosmic era to another can be
understood better with a variable $\omega$. So, in order to have a
continuous picture of cosmic evolution, $\omega$ should be
regarded as time-dependent. In the previous sections, by selecting
a simple power law expression of $t$ for the equation of state
parameter $\omega$, equivalence of the models $\Lambda\sim(\dot
a/a)^2$ and $\Lambda\sim\rho$ has been established in both four
and five dimensions. In the four dimensional case, the parameters
of the two equivalent models bear the same relationship with
cosmic matter and vacuum energy densities as was found by Ray et
al. \cite{Ray2007} for constant $\omega$. But in five dimensions,
the relationship of the same two parameters with $\Omega_m$ and
$\Omega_{\Lambda}$ differ from that of Pradhan et al.
\cite{Pradhan2005}. It has also been possible to show that the
sought-for signature flipping of the deceleration parameter $q$
can be obtained by suitable choice of $n$ in both four and five
dimensions. This feature of the work satisfies an important
criterion of $\Lambda$-CDM cosmology. It should be mentioned that
the model $\Lambda\sim(\ddot a/a)$ which was shown to be
equivalent to the present two $\Lambda$ models by Ray et al.
\cite{Ray2007} (in four dimensions) as well as by Pradhan et al.
\cite{Pradhan2005} (in five dimension) is found not to be
equivalent to the two models considered here when $\omega$ is
time-dependent.

If we now look at the plots, our observations regarding these are
as follows :\\ (1) The plots (Figs 1 and 2) for $\omega$ show that
in the early stage, it was positive i.e. in the early stages, the
Universe was matter-dominated. The plots for $q$ also support this
(i.e. decelerating phase of the Universe). At late times it is
evolving with negative values (i.e. at the present time). The
earlier real matter later on converted to the dark
energy-dominated phase of the Universe. It, finally, ends up at
$\omega = -1$, representing a cosmological constant dominated
Universe; \\ (2) The plots (Figs 3 - 6) for $q$ show that in the
early stages the Universe was in a decelerating phase and at late
times, it is in an accelerating one.

Finally, the models under consideration are shown to be equivalent
in both four and five dimensions and extension to the fifth
dimension do not present us any basically new physical feature
than the four dimensional case. So, a natural question can be
raised: is there any justification for extension to five
dimensions under the framework of general relativity? In this
connection we would like to mention the works of Kapner et al.
\cite{Kapner2007} and Rahaman et al. \cite{Rahaman2007} where it
is shown that extension of four dimensional general relativity to
higher dimensions becomes mostly ineffective.

\section*{Acknowledgments}
One of the authors (SR) is thankful to the authority of
Inter-University Centre for Astronomy and Astrophysics, Pune,
India for providing Visiting Associateship Programme under which a
part of this work was carried out.


                REFERENCES



\begin{thebibliography}{0}

\bibitem{Dunlop1996} J. Dunlop {\it et al.}, {\it Nature} {\bf 381}, 581 (1996).
\bibitem{Spinard1997} H. Spinard {\it et al.}, {\it Astrophys. J.} {\bf 484}, 581 (1997)
\bibitem{Riess1998} A. G. Riess {\it et al.}, {\it Astron. J.} {\bf 116}, 1009 (1998).
\bibitem{Perlmutter1999} S. J. Perlmutter {\it et al.}, {\it Astrophys. J.} {\bf 517}, 565 (1999).
\bibitem{Kujat2002} J. Kujat {\it et al.}, {\it Astrophys. J.} {\bf
572}, 1 (2002).
\bibitem{Bartelmann2005} M. Bartelmann {\it et al.}, {\it New Astron. Rev.} {\bf
49}, 199 (2005).
\bibitem{Chevron2000} S. V. Chevron and V. M. Zhuravlev, {\it Zh. Eksp. Teor. Fiz.} {\bf 118},
259 (2000).
\bibitem{Zhuravlev2001} V. M. Zhuravlev, {\it Zh. Eksp. Teor. Fiz.} {\bf 120},
1042 (2001).
\bibitem{Peebles2003} P. J. E. Peebles and B. Ratra, {\it Rev. Mod. Phys.} {\bf 75},
559 (2003).
\bibitem{Jimenez2003} R. Jimenez, {\it New Astron. Rev.} {\bf 47} 761
(2003).
\bibitem{Das2005} A. Das {\it et al.},  {\it Phy. Rev. D} {\bf 72},
043528 (2005).
\bibitem{Ratra1988} B. Ratra and P. J. E. Peebles, {\it Phy. Rev. D} {\bf 37},
3406 (1988).
\bibitem{Turner1997} M. S. Turner and M. White, {\it Phy. Rev. D} {\bf 56},
R4439 (1997).
\bibitem{Caldwell1998} Caldwell {\it et al.}, {\it Phy. Rev. Lett.} {\bf 80},
1582 (1998).
\bibitem{Liddle1999} A. R. Liddle and R. J. Scherrer, {\it Phy. Rev. D} {\bf 59},
023509 (1999).
\bibitem{Steinhardt1999} P. J. Steinhardt {\it et al.}, {\it Phy. Rev. D} {\bf 59},
123504 (1999).
\bibitem{Bhui2005} B. Bhui, B. C. Bhui and F. Rahaman, {\it Astrophys. Space.Sc.} {\bf 299},
61 (2005).
\bibitem{Rahaman2006} F. Rahaman, B. Bhui and B. C. Bhui, {\it Astrophys. Space.Sc.} {\bf 301},
47 (2006).
\bibitem{Rahaman2009} F. Rahaman, M. Kalam and S. Chakraborty, {\it Acta Phys. Polon. B} {\bf 40}, 25
(2009).
\bibitem{Freese1987} K. Freese {\it et al.}, {\it Nucl. Phys. B} {\bf 287},
797 (1987).
\bibitem{Ozer1987} M. {\"O}zer and M. O. Taha, {\it Nucl. Phys. B} {\bf 287},
776 (1987).
\bibitem{Chen1990} W. Chen and Y. S. Yu, {\it Phy. Rev. D} {\bf 41},
695 (1990).
\bibitem{Carvalho1992} J. C. Carvalho {\it et al.}, {\it Phy. Rev. D} {\bf 46},
2404 (1992).
\bibitem{Lima1994} J. A. Lima and J. M. F. Maia, {\it Phy. Rev. D} {\bf 49},
5597 (1994).
\bibitem{Ray2007} S. Ray, U. Mukhopadhyay and X. -H. Meng, {\it Grav. Cosmol.} {\bf 13},
142 (2007).
\bibitem{Pradhan2005} A. Pradhan {\it et al.}, {\it Int. J. Theor. Phys.} {\bf 47} 1751
(2008).
\bibitem{Mukhopadhyay2008} U. Mukhopadhyay, S. Ray and S. B. Duttachowdhury, {\it Int. J. Mod. Phys. D.}
{\bf 17}, 301 (2008).
\bibitem{Mukhopadhyay2009} U. Mukhopadhyay and S. Ray, {\it N. B. U. Math. J.} {\bf II}, 51 (2009).
\bibitem{Dymnikova1998} I. Dymnikova and M. Khlopov, {\it Grav. Cosmol. Suppl.} {\bf 4},
50 (1998).
\bibitem{Dymnikova2000} I. Dymnikova and M. Khlopov, {\it Mod. Phys. Lett. A} {\bf 15},
2305 (2000).
\bibitem{Dymnikova2001} I. Dymnikova and M. Khlopov, {\it Eur. Phys. J. C} {\bf 20},
139 (2001).
\bibitem{Huterer2001} D. Huterer and M. S. Turner, {\it Phys. Rev. D} {\bf 64},
123527 (2001).
\bibitem{Weller2002} J. Weller and A. Albrecht, {\it Phys. Rev. D} {\bf 65},
103512 (2002).
\bibitem{Polarski2001} D. Polarski and M.Chavellier, {\it Int. J. Mod. Phys. D} {\bf 10},
213 (2001).
\bibitem{Linder2003} E. V. Linder, {\it Phy. Rev. Lett.} {\bf 90},
91301 (2003).
\bibitem{Mukhopadhyay2007} U. Mukhopadhyay, P. P. Ghosh, M. Khlopov and S. Ray, gr-qc/07110686.
\bibitem{Knop2003} R. A. Knop {\it et al.}, {\it Astrophys. J.} {\bf 598}, 102 (2003)
\bibitem{Tegmark2004} M. Tegmark {\it et al.}, {\it Astrophys. J.} {\bf 606}, 702 (2004)
\bibitem{Sahni1999} V. Sahni {\it et al.}, {\it Pramana} {\bf 53}, 937 (1999)
\bibitem{Riess2001} A. G. Riess, {\it Astrophys. J.} {\bf 560}, 49 (2001).
\bibitem{Amendola2003} L. Amendola, {\it Mon. Not. R. Astron. Soc.} {\bf
342}, 221 (2003).
\bibitem{Padmanabhan2003} T. Padmanabhan and T. Roychowdhury,
{\it Mon. Not. R. Astron. Soc.} {\bf 344}, 823 (2003).
\bibitem{Kapner2007} D. J. Kapner {\it et al.}, {\it Phys. Rev. Lett.} {\bf 98}, 021101 (2007).
\bibitem{Rahaman2007} F. Rahaman {\it et al.}, {\it Int J Theor Phys} {\bf 48}, 3124 (2009).

\end{thebibliography}
\end{document}